\pdfoutput=1

\documentclass[aps,prl,reprint,groupedaddress]{revtex4-1}

\usepackage{graphicx}

\begin{document}

\title{Extended temporal association memory by inhibitory Hebbian learning }

\author{Tatsuya Haga}
\email{tatsuya.haga@riken.jp}
\author{Tomoki Fukai}
\email{tfukai@riken.jp}

\affiliation{RIKEN Center for Brain Science, Saitama, Japan}

\date{\today}

\begin{abstract}
Hebbian learning of excitatory synapses plays a central role in storing activity patterns in associative memory models. Furthermore, interstimulus Hebbian learning associates multiple items in the brain by converting temporal correlation to spatial correlation between attractors. However, growing experimental evidence suggests that learning of inhibitory synapses creates "inhibitory engrams", which presumably balance with the patterns encoded in the excitatory network. Controlling inhibitory engrams may modify the behavior of associative memory in neural networks, but the consequence of such control has not been theoretically understood. Noting that Hebbian learning of inhibitory synapses yields an anti-Hebbian effect, we show that the combination of Hebbian and anti-Hebbian learning can increase the span of temporal association between the correlated attractors. The balance of targetted and global inhibition regulates this span of association in the network. Our results suggest a nontrivial role of anti-Hebbian learning and inhibitory engrams in associative memory.
\end{abstract}

\maketitle


Animals can recall memory from imcomplete stimulus presentation; in other cases, presentation of one item leads to memory recall of a paired item. Such function is called associative memory. Hebb postulated that synchronous activation strengthens connections between neurons in the brain, and these strongly connected neuron ensembles (cell assemblies) are the basis of associative memory \cite{Hebb1949}. In the brain, this "Hebbian learning" is actually implemented as spike-timing-dependent plasticity (STDP) \cite{Bi1998,Mishra2016}. Furthermore, an attractor network model with Hebbian learning can recall activity patterns from incomplete external cues \cite{Hopfield1982}. Today, Hebb's postulate is a widely accepted paradigm for memory processing in the brain.

A number of experiments suggest that association between items are represented by correlations between activity patterns in the brain \cite{Deuker2016,Schapiro2016,Barron2017}. One important finding was made in the investigation of prolonged activity patterns in the temporal cortex of monkeys performing a visual working memory task \cite{Miyashita1988,Yakovlev1998}. After uncorrelated visual stimuli were consecutively presented during training, those stimuli evoked mutually correlated activity patterns in the test phase although the presentation order was random. Griniasty et al. proposed a model that bridges Hebbian learning and this finding \cite{Griniasty1993}. They added cross-stimulus terms to the local Hebbian connection matrix of the conventional associative memory model \cite{Hopfield1982}. The extended model converts the sequence of uncorrelated stimulus patterns into correlations between attractors, which are significantly correlated up to a separation of five in the sequence. Notably, this span of temporal association is robust against variations in model parameters and is consistent with experimental observations \cite{Griniasty1993,Amit1994}.

While Hebbian learning is sufficient for supporting the correlated attractors, the role of inhibitory learning for associative memory remains unclear. Actually, researchers are aware of the possible importance of inhibitory engrams in memory processing \cite{Barron2017}. Here, assuming that activity-dependent potentiation of inhibitory synapses effectively results in anti-Hebbian learning in associative memory models, we show that such learning induces previously unknown advantages in sequence coding with correlated attractors.

Let us assume a network of $N$ neurons. Below, $S_{i}=1,-1$ denotes activity of neuron $i$ (we will consider a 0/1 activity model later). Update of neural activity follows
\begin{eqnarray}
S_{i} (t+\delta t) = \mathrm{sign} \left[ \sum_{j=1}^{N} J_{ij} S_{j} (t) -\theta \right] ,
\label{eq:activity_model}
\end{eqnarray}
where $J_{ij}$ represents synaptic weights and $\theta$ is a threshold. The network stores $P$ random memory patterns $\xi_{i}^{\mu} \ (1 \leq i \leq N, 1 \leq \mu \leq P)$ that are biased as $\mathrm{E}[\xi_{i}^{\mu}]=a \ (-1 < a < 1)$ \cite{Amit1987}. We define synaptic weights as
\begin{eqnarray}
J_{ij} = \frac{1}{N} \sum_{\mu=1}^{P} ( c \hat{\xi}_{i}^{\mu} \hat{\xi}_{j}^{\mu} + \hat{\xi}_{i}^{\mu+1} \hat{\xi}_{j}^{\mu} + \hat{\xi}_{i}^{\mu} \hat{\xi}_{j}^{\mu+1} ) .
\end{eqnarray}
where $\hat{\xi}_{i}^{\mu}=\xi_{i}^{\mu}-a$ and $\xi_{i}^{P+1}=\xi_{i}^{1}$. The parameter $c$ can be either positive or negative. When $c$ is positive, this model is equivalent to that of Griniasty et al. \cite{Griniasty1993}. On the other hand, negative $c$ implies anti-Hebbian learning, which has not been extensively studied in associative memory.

We analyze attractors of this model following a similar procedure to the previous one \cite{Griniasty1993}. We define a pattern overlap, which represents the degree of coincidence between the instantaneous activity and the $\mu$-th memory pattern, as
\begin{eqnarray}
m^{\mu} = \frac{1}{N(1-a^2)} \sum_{j=1}^{N} \hat{\xi}_{i}^{\mu} S_{i}.
\end{eqnarray}
In the limit of $N \rightarrow \infty$, we can obtain the following mean-field equations from Eq. (\ref{eq:activity_model}):
\begin{eqnarray}
m^{\mu} = \bigg\langle \hspace{-5pt} \bigg\langle \hat{\xi}^{\mu} \mathrm{sign} \left[ \sum_{\alpha=1}^{P} ( c \hat{\xi}^{\alpha} + \hat{\xi}^{\alpha+1} + \hat{\xi}^{\alpha-1} ) m^{\alpha} \right] \bigg\rangle \hspace{-5pt}\bigg\rangle ,
\end{eqnarray}
where $\langle \hspace{-1pt} \langle \cdot \rangle \hspace{-1pt}\rangle$ denotes averaging over possible configurations of $\xi^{\mu}$. We calculate solutions (fixed points) to the simultaneous equations. The initial condition is $m^{\mu}=1$ for $\mu=\mu_{\mathrm{init}}$ and $m^{\mu}=0$ otherwise.  When the number of patterns is small, we can exactly calculate these solutions over all possible combinations of $\{ \xi^{\mu} \}$. However, when we increase the number of patterns, the number of possible configurations of $\xi^{\mu}$ (i.e., sublattices) rapidly diverges and becomes intractable. To overcome this difficulty, we perform the Monte-Carlo approximation of the mean-field equation by sampling a finite but large enough number of $\{ \xi^{\mu} \}$ ($10^6$ samples). Additionally, we calculate correlations between two attractors centered on patterns $\mu_\mathrm{init}$ and $\mu_\mathrm{init}+\nu$ as
\begin{eqnarray}
C(\nu) = \frac{1}{|C|} \bigg\langle \hspace{-5pt}  \bigg\langle \left( S(\mu_\mathrm{init}) - \bar{S} \right) \left( S(\mu_\mathrm{init}+\nu) - \bar{S} \right) \bigg\rangle \hspace{-5pt} \bigg\rangle ,
\end{eqnarray}
where
\begin{eqnarray}
|C| &=& 1-\bar{S}^2 , \\
S(\mu) &=& \mathrm{sign} \left[ \sum_{\alpha=1}^{P} \hat{\xi}^{\alpha} ( c m_{\mu}^{\alpha} + m_{\mu}^{\alpha+1} + m_{\mu}^{\alpha-1} )  \right] ,
\end{eqnarray}
and $m_{\mu}^{\rho}$ denotes the overlap between the pattern $\rho$ and the attractor retrieved from the memory pattern $\mu$. Assuming translation invariance, we calculated $S(\mu_\mathrm{init}+\nu)$ based on the attractor for $\mu_\mathrm{init}$. Mean activity $\bar{S}$ was calculated by
\begin{eqnarray}
\bar{S} = \bigg\langle \hspace{-5pt} \bigg\langle \mathrm{sign} \left[ \sum_{\alpha=1}^{P} ( c \hat{\xi}^{\alpha} + \hat{\xi}^{\alpha+1} + \hat{\xi}^{\alpha-1} ) m^{\alpha} \right] \bigg\rangle \hspace{-5pt} \bigg\rangle .
\end{eqnarray}
We performed these calculation by Python3, using Numpy and Scipy libraries (we share codes in http://github.com/TatsuyaHaga/antihebbhopfield).

\begin{figure}
\includegraphics{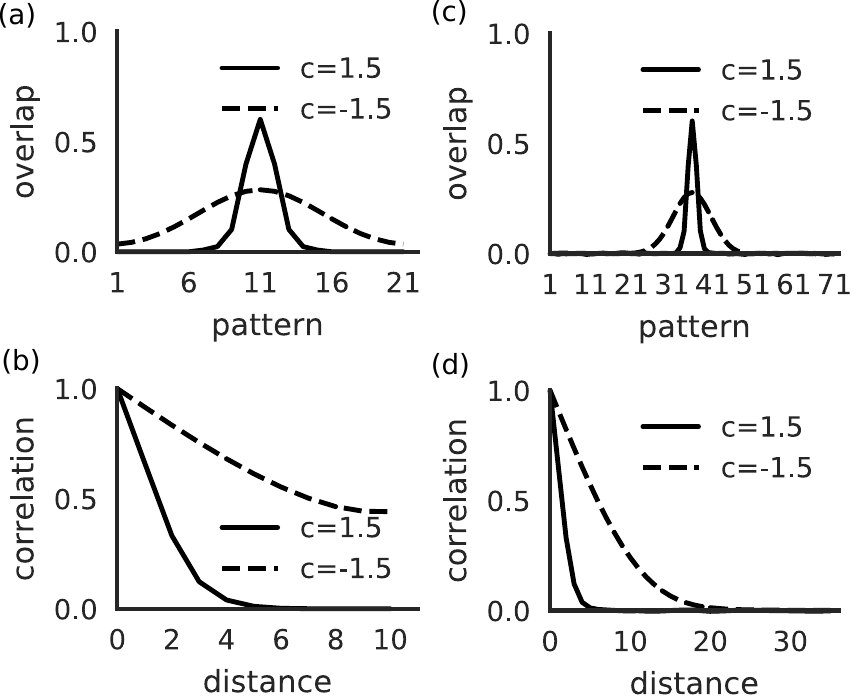}%
\caption{Expanded temporal association in anti-Hebbian learning. (a) Overlaps between a reference attractor ($\mu_{\mathrm{init}}=11$) and memory patterns. (b) Correlations between attractors. (c,d) Similar overlaps and correlations calculated with the Monte-Carlo approximation for $\mu_{\mathrm{init}}=36$. \label{fig1}}
\end{figure}

An unexpected finding is that negative $c$ significantly expands the span of temporal association among correlated attractors. Figure 1(a) and 1(b) shows solutions for $a=0, \theta=0$ and $P=21$  without the Monte-Carlo approximation. When $c$ is positive ($c=1.5$), our model reproduces the result shown by Griniasty et al. \cite{Griniasty1993} in which the neighboring attracters are significantly correlated up to the distance of five. In contrast, when $c$ is negative ($c=-1.5$), the correlation distance extends beyond 10. To see how the correlation behaves at longer distances, we obtained solutions for $P=71$ by using the Monte-Carlo approximation (Fig. 1(c) and 1(d)). The results show that the correlation between attractors extends up to the distance of 20, which is four times longer than that for $c=1.5$. 

\begin{figure}
\includegraphics{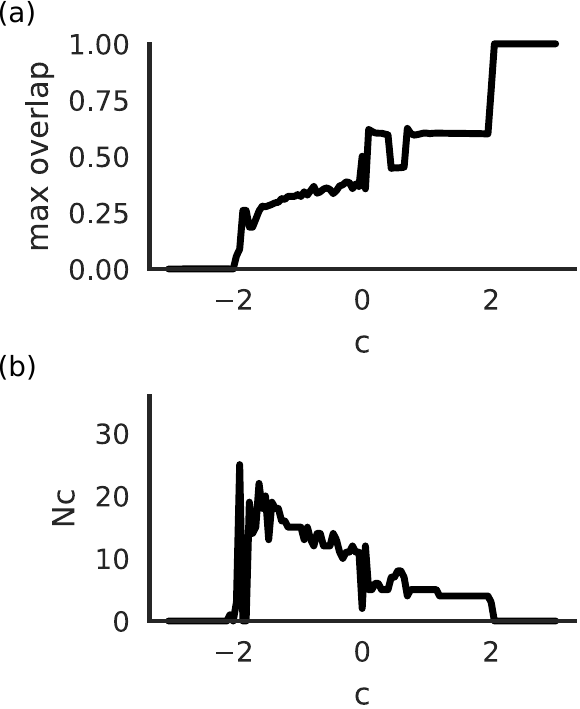}%
\caption{Parameter dependence of maximum overlaps (a) and $N_c$ (b) for unbiased stimulus patterns. \label{fig2}}
\end{figure}

Next, we quantitatively study how the value of $c$ changes attractors in our model by calculating the approximate solutions in the range $-3 \leq c \leq 3$ for $a=0, \theta=0$ and $P=71$. We calculated two measures: the maximum overlap that indicates successful memory retrieval, and the span of correlation $N_c$ defined as
\begin{eqnarray}
N_c=\mathrm{min} \{ \nu |  C(\nu)<10^{-2} \} -1 .
\end{eqnarray}
If only the nearest neighbour attractors have correlations greater than $10^{-2}$, $N_c$ is unity. The maximum overlap takes non-zero values only for $c>-2$ (Fig. 2(a)). The value of $N_c$ is robustly around five for $0<c<2$ (Fig. 2(b)) and becomes 0 for $c>2$ (that is, no correlated attractors exist in this range). Thus, for $c>0$ the threshold value $10^{-2}$ reproduces the results obtained by Griniasty et al. ($N_c=5$) \cite{Griniasty1993}. By contrast, as $c$ is decreased from 0 to -2, $N_c$ gradually increases even beyond 20 (Fig. 2(b)).

\begin{figure}
\includegraphics{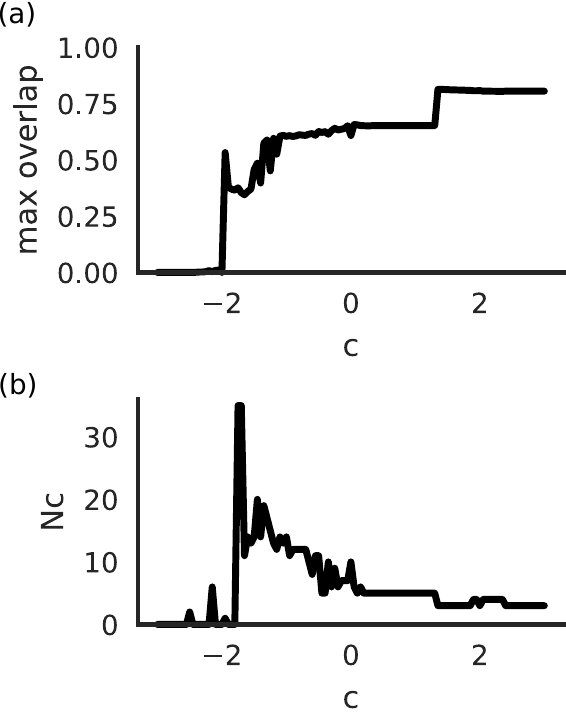}%
\caption{Parameter dependence of the maximum overlaps (a) and $N_c$ (b) for biased memory patterns. \label{fig3}}
\end{figure}

We can observe a similar expansion of correlation for biased patterns ($a=-0.8$, which corresponds to 10 \% activity level). The maximum overlap takes non-zero values for $c>-2$ (Fig. 3(a)), and $N_c$ increases as $c$ decreases in the negative value range (Fig. 3(b)). In sum, the extended span of correlation is generally found in the range $-2<c<0$ regardless of the bias of memory patterns.

\begin{figure}
\includegraphics{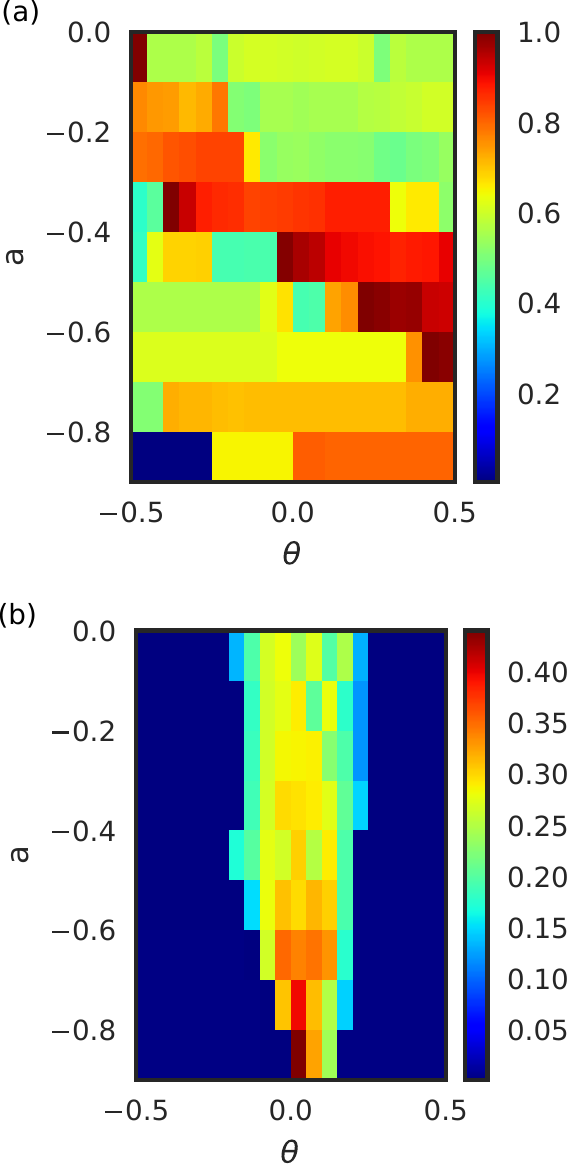}%
\caption{Maximum overlaps calculated under various parameter values (bias $a$ and threshold $\theta$). (a) $c=1.5$. (b) $c=-1.5$. \label{fig4}}
\end{figure}

We further examine the effect of firing threshold $\theta$ on the correlated attractors. Fig. 4 summarizes the maximum overlaps obtained in different settings of the bias $a$ and threshold $\theta$. In agreement with the previous report \cite{Griniasty1993}, when $c$ is positive ($c=1.5$) retrieval of a memory pattern occurs robustly in a broad region of the parameter space (Fig. 4(a)). In contrast, at $c=-1.5$ the model gives non-zero overlaps only in the vicinity of $\theta=0$ regardless of the value of $a$ (Fig. 4(b)). This result is qualitatively different from the conventional associative memory models in which storage capacity for biased patterns is optimized by a non-zero threshold \cite{PerezVicente1989}.

The class of associative memory models proposed here receives some support from recent findings of symmetric STDP and inhibitory engrams in the brain. To see this, we express neural activity and stored patterns by $V_{i}=0,1$ and $\eta_{i}^{\mu}=0,1$, with the mean activity $\mathrm{E}[\eta_{i}^{\mu}]=p$. Following the previous studies \cite{Buhmann1989,Tsodyks1988}, we set synaptic weights as
\begin{eqnarray}
J_{ij} = \frac{1}{N} \sum_{\mu=1}^{P} ( c \hat{\eta}_{i}^{\mu} \hat{\eta}_{j}^{\mu} + \hat{\eta}_{i}^{\mu+1} \hat{\eta}_{j}^{\mu} + \hat{\eta}_{i}^{\mu} \hat{\eta}_{j}^{\mu+1} ) , 
\end{eqnarray}
where $\hat{\eta}_{i}^{\mu}=\eta_{i}^{\mu}-p$. We can decompose this synaptic weight into excitation and inhibition as
\begin{eqnarray}
 J_{ij} = J_{ij}^{ \mathrm{E} }-J_{ij}^{ \mathrm{I} } ,
\end{eqnarray}
where
\begin{eqnarray}
 J_{ij}^{\mathrm{E}} &=& \frac{1}{N} \sum_{\mu=1}^{P} ( 2 \eta_{i}^{\mu} \eta_{j}^{\mu} + \eta_{i}^{\mu+1} \eta_{j}^{\mu} + \eta_{i}^{\mu} \eta_{j}^{\mu+1} + 4 p^2) \\ 
 J_{ij}^{\mathrm{I}} &=& (4-c') \frac{1}{N} \sum_{\mu=1}^{P} (\eta_{i}^{\mu} \eta_{j}^{\mu} + p^2) \nonumber \\
 && + c' \frac{p}{N} \sum_{\mu=1}^{P} (\eta_{i}^{\mu} + \eta_{j}^{\mu} ) ,
\end{eqnarray}
and $c'=c+2$. We note that $J_{ij}^{\mathrm{E}} \geq 0$ and $J_{ij}^{\mathrm{I}} \geq 0$ in the parameter region relevant to the phase transitions ($0 \leq c' \leq 4$). 

First, the excitatory weights involve terms symmetric with respect to $\eta^{\mu}$ and $\eta^{\mu+1}$. On the millisecond range timescale, these terms may emerge through a symmetric spike-timing-dependent plasticity with a broad time window. Actually, such a STDP rule has been recently revealed in the hippocampal area CA3 \cite{Mishra2016}. Alternatively, consecutive stimuli presented on a longer timescale can be correlated by the mechanism described previously \cite{Brunel1996, Yakovlev1998}. Second, the inhibitory weights consist of two terms: the first term represents anti-Hebbian learning (targeted inhibition) and the second term is global inhibition proportional to the total local activity of stored patterns. When $c'$ varies between 0 and 4, the balance of the two inhibition terms changes and so does the span of correlations between attractors. Because the targeted inhibition may correspond to inhibitory engrams \cite{Barron2017}, we propose that learning and control of inhibitory engrams regulates this balance to alter the dynamical behavior of correlated attractors in the brain, specifically in the hippocampus.

We can qualitatively understand why the model has broadly correlated attractors by means of energy function:
\begin{eqnarray}
E = - \frac{1}{N}  \sum_{i,j} J_{ij} S_{i} S_{j}.
\end{eqnarray}
We can rewrite the energy function in terms of pattern overlaps as
\begin{eqnarray}
E &=& - c \sum_{\mu=1}^{P} (m^{\mu})^2 - 2 \sum_{\mu=1}^{P} m^{\mu}m^{\mu+1} \nonumber \\
 &=& - c' \sum_{\mu=1}^{P} (m^{\mu})^2 + \sum_{\mu=1}^{P} (m^{\mu}-m^{\mu+1})^2 .
\end{eqnarray}
When $c'<0$, this function is trivially minimized when all overlaps vanish. Furthermore, if $c'=0$, there is no point minima because $E=\sum_{\mu=1}^{P} (m^{\mu}-m^{\mu+1})^2 $ is always zero when all overlaps take the same value. However, if $c'>0$ (i.e., $c>-2$), energy minimization requires the maximization of $(m^{\mu})^2$ under the penalty of $(m^{\mu}-m^{\mu+1})^2$. Without the penalty, the model is equivalent to the standard Hopfield model and generates a sharp peak of an overlap. However, the penalty term creates broadly distributed overlaps for small values of $c'$. As $c'$ increases, the relative contribution of the penalty becomes smaller, shrinking the distribution. 

We can obtain similar results through direct simulations of Eq. (\ref{eq:activity_model}) (data not shown) if we sequentially update neural activity or transmit temporally smoothed activities (slow synapses) \cite{Sompolinsky1986}. If we conduct a parallel update of all neurons, the model does not converge to stable states  when $c<0$. This sensitivity to updating methods is not seen in the conventional Hopfield-type models which practically behave similarly in sequential and parallel updates. Because the decrease of energy function of the Hopfield-type model is rigorously guaranteed only in sequential update \cite{Hopfield1982}, this difference may be due to shallower and more fragile landscape of energy function for $c<0$. It is intriguing to further clarify qualitative differences in retrieval dynamics between $c>0$ and $c<0$.

In sum, here we report that anti-Hebbian learning significantly expands the span of temporal association in associative memory models with correlated attractors. Our model predicts the nontrivial role of inhibitory engrams in regulating this effect, which may have significant implications for sequence memory encoding in the brain. 



\bibliography{ref.bib}

\end{document}